# Deterministic fabrication of large-area, high-crystallinity oxide moiré superlattices

*Reza Ghanbari[1†], Eli Rodrigues[1†], Young-Hoon Kim[2], Konnor Koons[1], Yan Li[3], Kabelo Lebogang[1], Yiming Ding[1], Doug Barefoot[1], Yueyin Wang[1], Yin Liu[1], Hua Zhou[4], Miaofang Chi[2,5], Ruijuan Xu[1]**

[1] Department of Materials Science and Engineering, North Carolina State University, Raleigh, NC, 27606, USA

[2] Center for Nanophase Materials Sciences (CNMS), Physical Sciences Directorate (PSD), Oak Ridge National Laboratory, Oak Ridge, TN, 37830, USA

[3] Materials Science Division, Advanced Photon Source, Argonne National Laboratory, Lemont, IL, 60439, USA

[4] X-ray Science Division, Advanced Photon Source, Argonne National Laboratory, Lemont, IL, 60439, USA

[5] Thomas Lord Department of Mechanical Engineering and Materials Science, Duke University, Durham, NC, 27708, USA

*Email: rxu22@ncsu.edu

[†] R. Ghanbari and E. Rodrigues contribute equally to this work.

**Abstract:** Oxide twistronics extends moiré engineering beyond van der Waals materials, offering a promising platform for accessing emergent interfacial phenomena arising from the strong coupling of lattice, charge, and orbital degrees of freedom in complex oxides. However, deterministic fabrication of high-crystallinity oxide moiré superlattices over large lateral dimensions remains challenging due to the three-dimensional bonding network of oxides. Here, we demonstrate a scalable, generalized fabrication strategy that enables the formation of high-crystallinity oxide moiré superlattices with clean, chemically bonded interfaces and precisely controlled twist angles down to nominal values of 0.1°, achieving sub-degree twist-angle accuracy across large contiguous lateral dimensions approaching the millimeter scale. Using $NaNbO_3$ as a model system, we show that the resulting interlayer coupling drives pronounced structural reconstruction that modifies both the phase structure and ferroelectric domain configuration. Synchrotron-based X-ray 3D reciprocal space mapping reveals the emergence of a single-phase state in twisted bilayers, in contrast to the mixed-phase structure observed in single-layer membranes prior to twist assembly. The structural signatures are further consistent with gradual lattice rotation distributed along the thickness direction that may accommodate interfacial shear strain, distinct from reconstruction observed in van der Waals moiré systems which primarily occurs through in-plane stacking rearrangement. This collective lattice response is correlated with twist-dependent nanoscale electromechanical modulations observed by piezoresponse force microscopy. These results establish a scalable materials platform for oxide twistronics and open new pathways towards integrating twist-engineered complex oxides into practical, macroscale device architectures.

## Introduction

By extending moiré engineering from van der Waals (vdW) materials to complex oxides, oxide twistronics has emerged as a new powerful platform in which strong interfacial coupling and intertwined lattice, charge, spin, and orbital degrees of freedom enable emergent interfacial phenomena beyond those accessible in conventional twisted vdW systems.[1-4] Recent experimental discoveries of topological polar vortices in twisted bilayer oxide membranes,[5, 6] together with theoretical predictions of flat bands, moiré magnetism, and correlated insulating states,[7-10] establish complex oxides as a new class of functional building blocks for moiré engineering. These exciting advances are driven by the development of crystalline freestanding oxide membranes,[11-17] which provide an oxide analogue to two-dimensional vdW materials while retaining a three-dimensional bonding network. By releasing epitaxial oxide thin films from their as-grown substrates through mechanical exfoliation or selective wet etching of a sacrificial buffer layer, the resulting transferable oxide membranes can be assembled with a controlled rotational mismatch at the interface, forming twist-engineered oxide moiré superlattices.[18, 19]

However, in contrast to vdW materials, the 3D bonding nature of complex oxides introduces fundamental fabrication challenges in realizing moiré superlattices with strongly coupled twist interfaces and precisely controlled twist angles, which are required for robust and rationally designed moiré phenomena.[20] First, oxide membrane surfaces are terminated with dangling covalent bonds that can adsorb hydroxyl species (i.e., $H_2O$ and OH groups) from wet etchant solutions or ambient air, as well as residue organic polymers incompletely removed after use as support layers during membrane etching and transfer. The presence of these surface contaminants hinders the formation of clean, chemically bonded interfaces. Second, oxide membranes are typically fabricated and assembled via wet-transfer processes that involve chemical etching of sacrificial buffer layers and transfer of the membranes in the presence of residual liquid. The liquid induces membrane motion and uncontrolled rotation during stacking, severely limiting the achievable precision in twist angle control.

To address these fundamental limitations, recent work has developed high-temperature thermal annealing using $CO_2$ laser to create atomically clean interfaces between symmetry-mismatched oxides, such as sapphire and $SrTiO_3$.[21] This method has subsequently been adapted to milder thermal conditions through furnace annealing, enabling the bonding between oxide membranes containing volatile elements (e.g., sodium) while preventing significant compositional loss of volatile elements during high-temperature annealing.[22] In parallel, a tear-and-stack method, widely used in the fabrication of 2D moiré superlattices, has recently been demonstrated for constructing oxide moiré superlattices with precise twist-angle control.[23] Despite these progresses, no existing approach has yet achieved high-crystallinity twisted oxide moiré superlattices over millimeter lateral dimensions with both clean interfaces and precise twist-angle control, capabilities that are both fundamentally essential for investigating robust moiré phenomena and technically imperative for advancing moiré twistronics towards scalable device deployment.

In this work, we demonstrate a scalable, generalized fabrication strategy that enables the formation of high-crystallinity oxide moiré superlattices with clean, chemically bonded interfaces and precisely controlled twist angles down to nominal values of 0.1º, achieving sub-degree twist-angle accuracy across large contiguous lateral dimensions approaching the millimeter scale. Leveraging high-crystalline-quality freestanding oxide membranes synthesized via atomic-scale thin film epitaxy as functional building blocks, we achieve deterministic twist assembly using pre-patterned alignment markers and precise rotational control enabled by a 2D materials transfer stage. Combined with our recently developed annealing protocol, this approach enables the formation of clean, chemically bonded twist interfaces between oxide membranes at optimized annealing temperatures, including systems containing volatile elements. Using $NaNbO_3$ as a model system, we demonstrate twisted bilayers with deterministically defined twist angles (e.g., nominal values of 0.1º, 7º, and 15º with measured deviations of only ~0.2–0.3°) and chemically bonded interface. The resulting interlayer coupling drives pronounced structural reconstruction in $NaNbO_3$ that modifies both the phase structure and ferroelectric domain configuration. Synchrotron-based X-ray 3D reciprocal space mapping (3D-RSM) reveals the emergence of a single-phase structural state in twisted bilayers, in contrast to the mixed-phase structure observed in single-layer membranes prior to twist assembly. Notably, we probe the structural signatures that are further consistent with gradual continuous lattice rotation distributed along the thickness direction that may accommodate interfacial shear strain, distinct from reconstruction in 2D vdW moiré systems, where strain accommodation occurs primarily through in-plane stacking rearrangement. The resulting collective lattice response is correlated with twist-dependent nanoscale electromechanical modulations observed by piezoresponse force microscopy (PFM). Overall, our results establish a scalable fabrication workflow for oxide twistronics, providing a robust platform that opens new opportunities to advance twist-engineered complex oxides toward practical device integration and macroscale functional deployment.

## Results and Discussion

Epitaxial heterostructures consisting of 5-15 nm thick oxide films were synthesized on single-crystalline $SrTiO_3$ substrates with a sacrificial epitaxial buffer layer ($La_{0.7}Sr_{0.3}MnO_3$ or $Sr_{1.5}Ca_{1.5}Al_2O_6$) via pulsed laser deposition. To enable large-area, millimeter-lateral-scale oxide moiré superlattices with atomically coherent interfaces and precisely controlled twist angles, we develop a controlled twist-assembly protocol that integrates a deterministic alignment strategy with systematically optimized annealing treatments (Figure 1a; Methods). Specifically, arrays of cross-shaped chromium markers parallel to the [100] and [010] crystallographic directions were fabricated on the film surface prior to lift-off (Figure 1b), serving as optical alignment references for stacking bilayer membranes with a controlled rotational offset. Next, the as-grown oxide heterostructures were diced into four square pieces (~ 2 mm × 2 mm), serving as millimeter-scale building blocks for constructing twisted heterostructures. The sacrificial buffer layer was then selectively dissolved in either deionized water or a dilute acidic solution, releasing the freestanding oxide membrane. The first membrane layer, supported by a spin-coated polymethyl methacrylate (PMMA) layer, was transferred onto a destination substrate using a conventional wet-transfer method via scooping with a metal loop, forming the bottom layer of the twisted moiré heterostructure. After removing PMMA in acetone, the membrane was annealed in a tube furnace in flowing oxygen to obtain a chemically clean surface prior to stacking assembly. The second membrane layer was released using a bilayer polymer support consisting of PMMA and polydimethylsiloxane (PDMS), where the PDMS enables deterministic pickup and controlled release of the membrane at elevated temperature, allowing precise placement during stacking. Twisted bilayer assembly was conducted using a commercial motorized 2D materials transfer stage equipped with an optical microscope, a heating sample stage, and a transfer arm with translational ($x$, $y$, $z$) and rotational degrees of freedom, with angular resolution down to 0.001°. After aligning the top and bottom layers through optically resolvable alignment markers, the top membrane was rotated to the desired twist angle and brought into contact with the bottom membrane to form a controlled twist assembly. Subsequent heating weakens the adhesion between PDMS and PMMA, enabling release of the bilayer stack from the PDMS support. Following PMMA removal, we applied an annealing protocol optimized for complex oxide membranes, including those containing volatile elements,[22] to promote interfacial chemical bonding while minimizing compositional degradation. Using this approach, we fabricated twisted moiré heterostructures with large contiguous lateral dimensions approaching the millimeter scale across a variety of complex oxides, including ferroelectric $NaNbO_3$ (Figure 1c, d) and quantum paraelectric $SrTiO_3$ (Figure S1a) and $NaTaO_3$[24] (Figure S1b), with intended twist angles spanning 0.1° to 15°. Despite the mechanical fragility of ultrathin membranes, which often leads to minor localized cracks or edge imperfections during transfer and assembly, the fabricated membranes retain large contiguous regions, suitable for macroscale characterization and potential device integration.

Next, we evaluate the precision of twist angle control in the fabricated moiré heterostructures through atomic-resolution scanning transmission electron microscopy (STEM) characterization, taking high-crystallinity twisted $NaNbO_3$ bilayers as a model system. The heterostructures consist of two 6 nm-thick membranes (~ 15 unit cells each) assembled with nominal twist angles of 0.1º, 7º, and 15º. Atomic-resolution high-angle annular dark-field (HAADF) STEM imaging, combined with fast Fourier transform (FFT) analysis, was performed on bilayer membranes transferred onto TEM grids (Figure 2a-c and Figure S1c, d). Plan-view STEM images clearly resolve moiré superlattices with periodicities that vary systematically with twist angle. FFT analysis further reveals two distinct sets of diffraction patterns corresponding to the individual membrane layers, where the angular offset between these diffraction patterns directly quantifies the relative in-plane twist angle. Across all samples, the measured twist angles closely match the intended nominal values, with deviations within 0.2–0.3°, demonstrating deterministic twist-angle control at the atomic scale (Figure 2d). Furthermore, the experimentally extracted moiré periodicities from the HAADF images well agree with theoretical predictions for the corresponding twist angles. Specifically, we measure superlattice periodicities of approximately 80, 3, and 1.5 nm for the intended twist angles of 0.1°, 7°, and 15°, respectively, in close agreement with the theoretically predicted values of 74, 3.4, and 1.4 nm[23, 25, 26] (Figure 2e).

We then verify the clean, chemically bonded interfaces following the systematically optimized annealing protocol as part of our fabrication workflow. In the as-stacked state, cross-sectional STEM imaging reveals a ~ 2 nm amorphous interfacial layer arising from residual carbonaceous species introduced during polymer-assisted transfer, consistent with our previous studies[22] (Figure 3a). To restore interlayer bonding and enable robust moiré coupling, we implemented an oxygen annealing protocol in a conventional tube furnace. The annealing conditions were systematically optimized by varying temperature between 500–800 °C, while closely monitoring surface morphology evolution via atomic force microscopy (AFM) and compositional stability through optical contrast, which is particularly critical for detecting elemental loss in volatile element-containing oxides such as $NaNbO_3$. Using 15º twisted $NaNbO_3$ bilayers as a representative system, AFM measurements show a monotonic decrease in root-mean-square (RMS) roughness with increasing annealing temperature, from 996 pm in the as-stacked state to 158 pm after annealing at 660 °C with negligible additional improvement upon extended annealing beyond two hours (Figure 3c-e). At the optimal annealing temperature, two distinct sets of step terraces emerge, indicating interfacial reconstruction and restored crystalline continuity (Figure 3d). Annular dark-field (ADF) imaging further confirms an atomically abrupt interface, indicating complete removal of the amorphous interlayer and re-establishment of atomic coherence in the annealed heterostructures, leading to the formation of a chemically bonded perovskite network across the interface with continuous A-site and B-site stacking (Figure 3b). Detailed atomic-scale characterization of this bonding environment has been reported in our recent work.[22] Notably, the optimal annealing temperature is material-dependent (Figure S2). Following the same protocol, we identify optimal annealing temperatures of 700 ºC for $SrTiO_3$ and 600 ºC for $NaTaO_3$.

This fabrication workflow enables oxide moiré superlattices with large contiguous lateral dimensions approaching the millimeter scale and chemically bonded interfaces, providing a robust platform to probe intrinsic moiré phenomena in complex oxides. To further elucidate the structural evolution in precisely twist-controlled $NaNbO_3$ moiré heterostructures, we performed synchrotron X-ray 3D-RSM, to systematically investigate the strain state and phase structure of (i) the as-grown 15 nm thick $NaNbO_3$/$SrTiO_3$ epitaxial heterostructure, (ii) a released 15 nm single-layer membrane transferred onto silicon, and (iii) twisted bilayer moiré heterostructures on silicon with nominal twist angles of 0.1° and 1°, and each layer 15 nm thick. Our prior work reveals a strain-induced morphotropic phase boundary in $NaNbO_3/SrTiO_3$ heterostructures, characterized by coexisting monoclinically distorted $M_B$ and $M_C$ phases.[27] Consistent with these results, synchrotron RSM near the 002-diffraction clearly reveals $M_B$-$M_C$ phase coexistence in the coherently strained as-grown heterostructure, with out-of-plane lattice parameters of 3.884 Å ($M_B$) and 3.954 Å ($M_C$), respectively, consistent with our previous studies (Figure 4a, e). Upon release and transfer onto silicon, both the single-layer membrane and twisted bilayer membranes retain high crystalline quality, as evidenced by well-defined Laue fringes extended far along the Bragg rod (Figure 4b-d, f-h). Notably, phase coexistence persists in the released single-layer $NaNbO_3$ membranes, as indicated by peak splitting in the RSM near the 002-diffraction (Figure 4b, f). The monoclinic symmetry of the coexisting phases is further confirmed by characteristic two-fold ($M_B$) and three-fold ($M_C$) peak splitting in the HL plane near the 112-diffraction (Figure S3f). Compared to the strained as-grown heterostructure, the single-layer membrane exhibits partial strain relaxation, primarily reflected in $M_C$ phase with the out-of-plane lattice decreasing from 3.954 Å to 3.92 Å.

In contrast to the mixed-phase state observed in both the as-grown heterostructure and the released single-layer membrane, the twisted bilayer membranes exhibit a single-phase state, showing the absence of peak splitting in RSMs near 002-diffraction, independent of twist angles (Figure 4b- d, f- h). Further analysis near the 112-diffraction condition shows that the in-plane lattice parameters of both the 0.1° and 1° twisted heterostructures remain nearly identical to those of the relaxed $M_C$ phase in the single-layer membrane, indicating the absence of additional macroscale in-plane strain (Figure S3d- l). However, a pronounced azimuthal peak splitting is observed in the in-plane HK maps near the 112-diffraction for both twist angles (Figure 4m, n and Figure S4), consistent with the imposed in-plane twisting geometry. The 1° twisted bilayer exhibits a larger peak separation than the 0.1° bilayer, in agreement with its greater nominal twist angle (Figure 4i- l). Strikingly, instead of two discrete peaks corresponding to the two stacked membrane layers, we observe a continuous azimuthal peak distribution. This feature is unlikely to originate from non-uniform twisting of the two membranes, as STEM measurements acquired across multiple regions of the sample reveal a highly uniform twist angle, with variations as small as 0.06° (Figure 2d). We also note that alternative mechanisms such as twist inhomogeneity, bending, or mosaicity are not strongly supported by the experimental data, given the absence of asymmetric peak broadening or diffuse scattering. The observed continuous azimuthal distribution is therefore consistent with a gradual in-plane lattice rotation distributed along the thickness direction, rather than an abrupt rotational discontinuity of the twist angle localized solely at the twist interface. These observations suggest that the interfacial misorientation may propagate across the heterostructure thickness, with lattice layers near the interface undergoing progressive rotational distortion to accommodate the imposed twist. Such a distributed shear distortion along the out-of-plane direction could provide an efficient mechanism for relieving interfacial shear strain without maintaining a sharp rotational mismatch at the interface. These observations further support the formation of a chemically bonded interface that influences the phase structure and suggest that the strong covalent bonding network in complex oxides may promote structural reconstruction through collective lattice distortion along the out-of-plane direction. Due to the differences in bonding between oxides and vdW materials, this behavior appears fundamentally distinct from that typically observed in 2D vdW moiré superlattices, where weak interlayer coupling largely confines strain accommodation to the interface, often resulting in modified stacking configurations such as triangular AB/BA domains or soliton networks.[25, 28-31] In contrast, the three-dimensional ionic-covalent bonding network in complex oxides may enable interfacial misorientation to be distributed throughout the film thickness, potentially suppressing localized strain accumulation and promoting structural coherence. The stability of the observed lattice rotation behavior may be understood in terms of a critical thickness governed by the competition between elastic and interfacial energies. In twisted oxide membranes, the imposed misorientation could in principle be accommodated either through a continuous lattice rotation distributed across the film thickness or through localization of the misorientation at the interface. For sufficiently thin membranes, the elastic energy cost associated with distributed lattice rotation may remain relatively small, allowing a coherently strained state to be stabilized. As thickness increases, the accumulated elastic penalty may favor localization of the misorientation near the interface. The critical thickness would therefore correspond to the crossover between these competing mechanisms. Further detailed, depth-resolved atomic-scale STEM studies will be necessary to directly resolve this proposed lattice rotation behavior, quantify the critical thickness, and elucidate its dependence on bonding strength, lattice symmetry, and elastic anisotropy across different complex oxides.

To understand how the reconstructed twist interface influences ferroelectric domains in $NaNbO_3$, we next examine its impact on domain configuration using PFM. In contrast to the RSM measurements, where we focused on 15 nm thick single-layer membranes and their corresponding twisted bilayers, the PFM measurements were performed on significantly thinner membranes to enhance interfacial contributions: 5 nm thick single-layer and 12 nm thick twisted bilayers consisting of 6 nm membranes with intended twist angles of 0.1º and 1º (Figure 5 and Figure S5). Since PFM is a surface-sensitive technique that probes the local piezoresponse near the membrane surface, reducing the total thickness increases the relative contribution of twist-interface-driven structural modulations. PFM measurements show that the single-layer $NaNbO_3$ membrane transferred onto silicon exhibits a pronounced in-plane ferroelectric response with no detectable out-of-plane polarization component. This behavior arises from the strong depolarization field in ultrathin freestanding membranes, which suppresses out-of-plane polarization.[32] The absence of out-of-plane polarizations further stabilizes a distorted Mc phase at this thickness with polarizations along the in-plane [100] or [010] directions (Figure 5a).

For the 0.1° and 1° twisted bilayers, the large-scale domain morphologies remain similar to those of the single-layer membrane, as highlighted by the blue outlines (Figure 5a–c). However, higher-magnification imaging reveals clear local modifications of the ferroelectric response that are reminiscent of moiré-coupled features. In the 0.1° twisted bilayer, periodic modulations appear within the ferroelectric domains, with a characteristic length scale of approximately 40 nm (Figure 5d, e), consistent with the theoretically predicted and experimentally confirmed moiré periodicity for this twist angle (Figure 2a). These periodic features correlate directly with the moiré superlattice geometry, suggesting that the local electromechanical response is influenced by spatially varying interlayer registry and the associated lattice distortions. While the observed PFM response raises the possibility of complex polarization textures, such as vortex-like features as suggested in previous studies[5], this level of structural detail lies beyond the spatial resolution of PFM and direct identification of such polar topologies would require higher-resolution techniques, such as atomic-resolution STEM or emerging approaches including multislice electron ptychography capable of resolving atomic-scale polarization displacements arising from anion displacements in $NaNbO_3$.[33] It is important to note that PFM probes the electromechanical response integrated across the membrane thickness, including contributions from both the reconstructed twist interface and regions away from the interface. Because shear-strain-induced lattice modulation likely relaxes gradually away from the interface, the moiré-induced contrast appears less spatially uniform in PFM measurements (Figure S6) than in plan-view STEM imaging (Figure 2a-c). For the 1° twisted bilayer, the expected moiré periodicity (~10 nm, Figure 5f) falls below the lateral resolution limit of PFM (~ 20 nm). Instead of a well-resolved long-range periodic modulation, the ferroelectric response exhibits localized clustered contrasts embedded within a large-domain background (Figure 5c and Figure S7). The clear differences between the single-layer and twisted bilayers indicate that the reconstructed twist interface, arising from strong interlayer coupling, modifies the local polarization configuration and influences the overall electromechanical response and ferroelectric domain structure. These results establish PFM as a complementary and functionally sensitive probe of oxide moiré superlattices. Unlike plan-view STEM, which can conflate true interfacial coupling with simple projection of two uncoupled rotated lattices, PFM directly measures local electromechanical response, providing additional evidence that the reconstructed interface influences the polarization-related responses in twisted oxide heterostructures.

**Conclusion**

This work establishes a scalable fabrication strategy for constructing large-area oxide moiré superlattices with deterministic twist-angle control and clean, chemically bonded interfaces, broadly applicable across a wide range of complex oxide materials. By integrating precise layer stacking with material-specific annealing protocols, we overcome key processing constraints inherent to complex oxides and achieve twist-angle control down to nominal values of 0.1º, achieving sub-degree twist-angle accuracy across large contiguous lateral dimensions approaching the millimeter scale. Structural and functional characterizations support the presence of strong interlayer coupling in these oxide moiré superlattices, driving pronounced structural reconstruction that influences both the phase structure and ferroelectric domain configuration**.** Notably, rather than exhibiting behavior fully consistent with an abrupt rotational discontinuity localized at the twist interface, the oxide moiré system shows structural signatures consistent with gradual in-plane lattice rotation distributed along the out-of-plane thickness direction, which may provide a mechanism for accommodating interfacial shear strain. This behavior appears distinct from reconstruction in 2D vdW moiré systems, where strain accommodation occurs predominantly through in-plane stacking rearrangements. The resulting collective lattice response is correlated with twist-dependent nanoscale electromechanical modulations observed by PFM. These findings demonstrate that the controlled twist degree of freedom provides an effective parameter for engineering phase structure and polarization response in complex oxides. More broadly, our results shift the focus of oxide moiré research beyond electron microscopy–based assessments of interfacial alignment toward understanding how chemically bonded interfaces influence structural reconstruction and emergent functional behavior in oxide moiré superlattices. Ultimately, the realization of oxide moiré superlattices with large contiguous lateral dimensions offers a pathway for integrating twist-engineered functionalities into practical device architectures. In particular, such scalability may enable macroscale capacitor-based and transport measurements of emergent moiré phenomena, the development of optical metasurfaces, and integration with existing semiconductor technologies, helping bridge oxide twistronics from fundamental studies toward scalable device platforms.

## Materials and Methods

### $NaNbO_3$ thin film synthesis

Epitaxial heterostructures consisting of 6- and 15-nm-thick $NaNbO_3$ films and 20-nm-thick $La_{0.7}Sr_{0.3}MnO_3$ (LSMO) sacrificial layers were synthesized on (001)-oriented single-crystal $SrTiO_3$ substrates by pulsed-laser deposition. The LSMO sacrificial layer was deposited at 730 °C in a dynamic oxygen pressure of 200 mTorr, using a laser fluence of 1.7 J $cm^{-2}$ and a repetition rate of 3 Hz, with a 3.7 $mm^2$ imaged laser spot. $NaNbO_3$ films were grown at 660 °C in a 210 mTorr of dynamic oxygen pressure, a laser fluence of 2.1 J $cm^{-2}$, and a repetition rate of 2 Hz, with a 4.58 $mm^2$ imaged laser spot. The heterostructures were subsequently cooled down to room temperature at a rate of 5 °C $min^{-1}$ in a static oxygen pressure of 2.5 Torr.

### $SrTiO_3$ thin film synthesis

Epitaxial heterostructures consisting of 5-nm-thick $SrTiO_3$ films and 15-nm-thick $Sr_{1.5}Ca_{1.5}Al_2O_6$ sacrificial layers were synthesized on (001)-oriented single-crystal $SrTiO_3$ substrates using pulsed laser deposition. The $Sr_{1.5}Ca_{1.5}Al_2O_6$ sacrificial layer was deposited at a heater temperature of 900 °C, a dynamic oxygen pressure of 75 mTorr, a laser fluence of 1.7 J $cm^{-2}$, a repetition rate of 1 Hz, and an imaged spot size of 4.1 $mm^2$. The STO film was deposited at a heater temperature of 760 °C, a dynamic oxygen pressure of 100 mTorr, a laser fluence of 1.2 J $cm^{-2}$, a repetition rate of 1 Hz, and a spot size of 2.6 $mm^2$. After deposition the heterostructures were cooled to room temperature at a rate of 10 °C $min^{-1}$ in a static oxygen pressure of 1.5 Torr.

### $NaTaO_3$ thin film synthesis

Epitaxial heterostructures consisting of 10-nm-thick $NaTaO_3$ films and 40-nm-thick LSMO sacrificial layers were synthesized on (001)-oriented single-crystal $SrTiO_3$ substrates using pulsed laser deposition. The LSMO sacrificial layer was deposited at a heater temperature of 730 °C, a laser fluence of 1.69 J $cm^{-2}$, with a repetition rate of 3 Hz, and imaged laser spot size of 3.7 $mm^2$, in a dynamic oxygen pressure of 210 mTorr. The $NaTaO_3$ films were deposited at a heater temperature of 660 °C, a laser fluence of 1.69 J $cm^{-2}$, a repetition rate of 2 Hz, and a laser spot size of 3.7 $mm^2$, in a dynamic oxygen pressure of 210 mTorr. The heterostructures were subsequently cooled to room temperature at a rate of 5 °C $min^{-1}$ in a static oxygen pressure of 2.45 Torr.

**Preparing samples for etching and twist assembly**

Following synthesis, 20-nm-thick chromium alignment markers were deposited onto the heterostructures by thermal evaporation (MBraun MB-EVAP) at a rate of $1.5 \times 10^{-2}$ Å $s^{-1}$, using either patterned photoresist or a shadow mask. After deposition, the photoresist was removed by acetone lift-off, leaving chromium alignment markers on the film surface. The samples were then diced into four square pieces (~ $2 \times 2$ $mm^2$) using a precision diamond saw (Leco VC-50) and sequentially cleaned by sonication in acetone and isopropyl alcohol (IPA), followed by drying using dry nitrogen. A PMMA (MicroChem PMMA A8) support layer was spin-coated at 10,000 rpm for 35 seconds with a ramp rate of 2,500 rpm $s^{-1}$ and baked on a hot plate at 135 °C for 3 minutes. The sample edges were then lightly scratched with sharp tweezers to expose the sacrificial layer and facilitate uniform etching.

**Membrane lift-off and transfer**

The samples were immersed in an etchant solution to remove the sacrificial layer. For $Sr_3Al_2O_6$, deionized (DI) water was used as the etchant. For $La_{0.7}Sr_{0.3}MnO_3$, a mixed etchant consisting of 50 ml DI water, 0.5 ml diluted HCl (prepared by diluting 37% HCl 1:1 with water), and 5 g KI was used. Following complete etching, a $SiO_2$/Si substrate was cleaned by sequential sonication in acetone and IPA, followed by oxygen plasma cleaning at medium intensity for five minutes using a Harrick Plasma PDC-32G plasma cleaner. The freestanding membrane was then transferred onto the cleaned $SiO_2$/Si substrate, serving as the first layer of the twisted bilayer heterostructure. The sample was heated on a hot plate at 90 °C for 15 minutes to promote adhesion, after which the PMMA layer was removed by immersion in acetone heated at 105 °C for 40 minutes, followed by a 3-minute IPA rinse.

**Stacking membranes into a twisted heterostructure**

The second membrane layer was transferred using a bilayer polymer support stack consisting of PDMS and PMMA. After etching the sacrificial buffer layer, the PDMS/PMMA-supported membrane, adhered onto a glass slide, was mounted onto the transfer arm of an all-electric 2D materials transfer stage (2DTrans-ams-03, Jooin Technology). The silicon substrate containing the first membrane layer was mounted on the bottom stage. Using precise control of the $x$, $y$, $z$ translational axes and the rotational $\theta$ axis, the membranes were first aligned at 0º twist angle using the chromium alignment markers as references. The desired twist angle was then applied by rotating the transfer arm. The top membrane was subsequently lifted to allow a drop of DI water to be introduced onto the bottom membrane, after which the membranes were laminated together. The bottom stage was heated to 90 °C for 20 minutes to enhance interfacial adhesion. The transfer arm was then carefully retracted, removing the PDMS layer and leaving the PMMA-supported membrane on the substrate, thereby forming the twisted bilayer heterostructure. Finally, the PMMA layer was removed using the same procedure described above. The twisted heterostructures were annealed in a tube furnace under flowing oxygen at 660 °C for two hours for $NaNbO_3$, 700 °C for two hours for $SrTiO_3$, and 600 °C for two hours for $NaTaO_3$.

**Synchrotron X-ray diffraction**

Synchrotron X-ray 3D reciprocal space mapping (3D-RSM) measurement was carried out on a six-circle Huber diffractometer configured with Psi-C geometry at the beamline 33-BM of the Advanced Photon Source at Argonne National Laboratory, utilizing X-rays with energies of 20 KeV (wavelength $\lambda = 0.6199$ Å). A Si (111) double-crystal monochromator with an energy resolution of $\Delta E/E = 1 \times 10^{-4}$ was employed to select the X-ray energy. The X-ray beam at 33-BM has a total flux of $5 \times 10^{11}$ photons per second at 20 keV with a slit-defined beam profile of 400 µm (Vertical) × 500 µm (Horizontal). Scattering patterns were acquired using a Pilatus 2D area detector, and the 2D images were subsequently processed. Geometric corrections were applied to all 3D-RSM data, which was further analyzed using the RSMap3D software developed at APS.

**Piezoresponse force microscopy**

Dual AC Resonance Tracking (DART) piezoresponse force microscopy was performed to scan the ferroelectric domains with an MFP-3D Origin+ AFM (Asylum Research) using a conductive Pt/Ir coated probe (Nanosensor, PPP-EFM, force constant ≈ 2.8 N $m^{-1}$).

**Scanning transmission electron microscopy**

To investigate the moiré superlattice, high-angle annular dark-field (HAADF) imaging was performed using an aberration-corrected STEM instrument (NeoARM200CF, JEOL, Japan) operated at an accelerating voltage of 200 kV. The probe convergence semi-angle and detector collection semi-angle were set to approximately 28 mrad and 68-270 mrad, respectively. Cross-sectional TEM specimens were prepared using a focused ion beam (FIB) system (Helios G4 UC DualBeam System, Thermo Fisher Scientific, USA).

**Data availability**

All data supporting the findings of this study are available within the Article and its Supplementary Information.

**Supporting Information**

Supplementary Figures 1-7.

**Acknowledgments**

We thank Aarushi Khandelwal and Prof. Harold Y. Hwang for their assistance with preliminary XRD characterizations and insightful discussions. R.G., K.L., Y.D. and R.X. acknowledge support from the National Science Foundation (NSF) under award No. DMR-2442399 and the American Chemical Society Petroleum Research Fund under award No. 68244-DNI10. K.K. and R.X. acknowledge support from the Army Research Office under award No. W911NF-25-1-0201. E.R. and D.B. acknowledge support from the NC State College of Engineering (COE) Research Experience for Undergraduate (REU) program, funded by the COE Enhancement Fee. This work was supported by the U.S. Department of Energy, Office of Basic Energy Sciences, Division of Materials Sciences and Engineering. Microscopy research conducted as part of a user project at the Center for Nanophase Materials Sciences (CNMS), which is a US Department of Energy, Office of Science User Facility at Oak Ridge National Laboratory. This research was performed on APS beam time award (https://doi.org/10.46936/APS-191841/60015614) from the Advanced Photon Source, a U.S. Department of Energy (DOE) Office of Science user facility operated for the DOE Office of Science by Argonne National Laboratory under Contract No. DE-AC02-06CH11357. Y.W. and Y.L. acknowledge the support by the U.S. National Science Foundation (NSF) under Award No. DMR-2340751.

## Main Figures

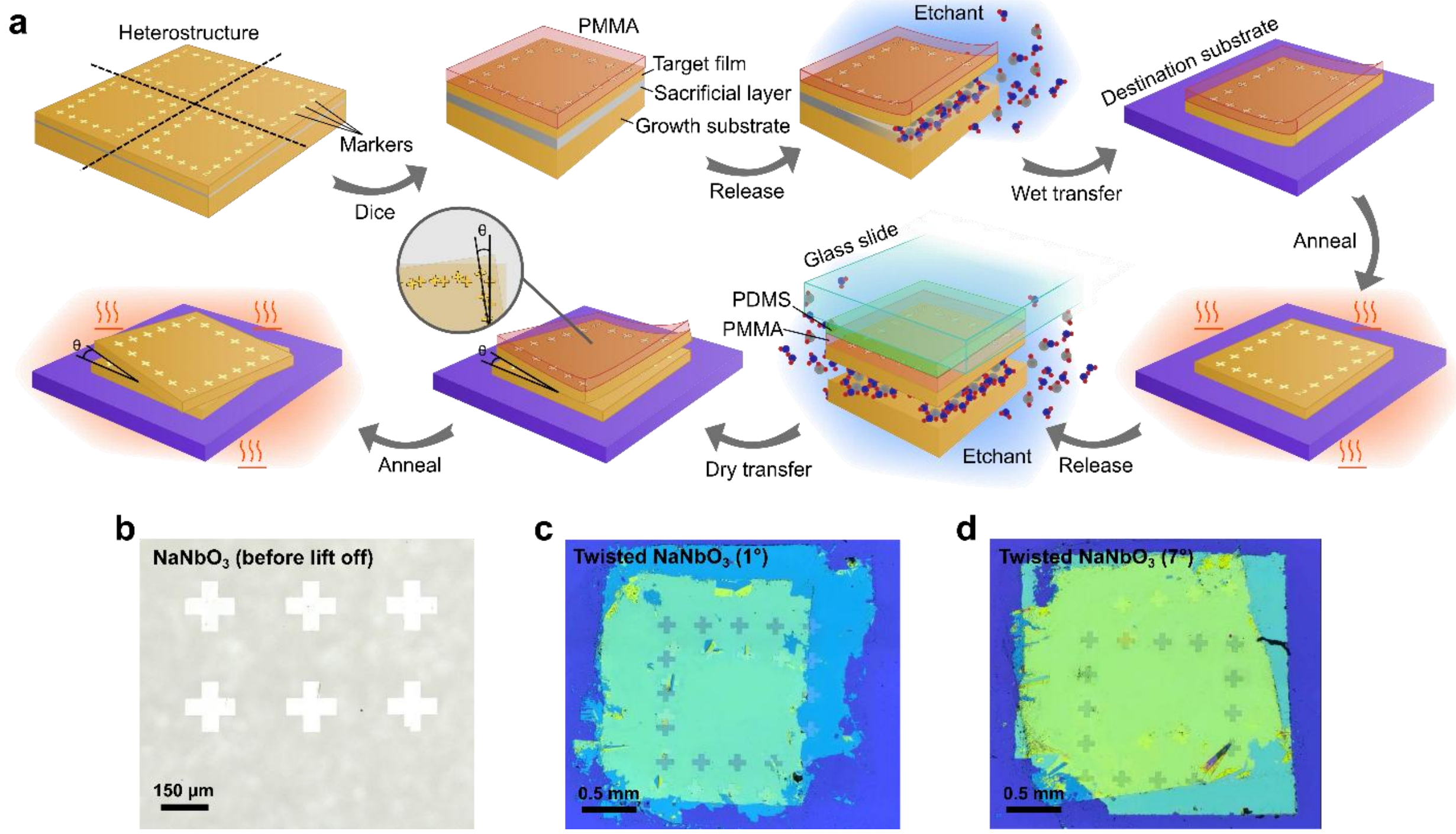

**Figure 1. Fabrication of twisted bilayer oxide membranes with large contiguous lateral dimensions approaching the millimeter scale, precise twist angle control, and clean interfaces.** **(a)** Schematic illustration of the fabrication workflow for twisted bilayer oxide membranes, incorporating chromium alignment markers and post-synthesis thermal annealing to achieve deterministic twist angles and a clean twist interface. Optical images of **(b)** patterned chromium alignment markers on as-grown $NaNbO_3/La_{0.7}Sr_{0.3}MnO_3/SrTiO_3$ (001) heterostructures, and $NaNbO_3$ bilayer membranes with **(c)** a 1° twist angle (6 nm each layer) and **(d)** a 7° twist angle (15 nm each layer) on $SiO_2/Si$.

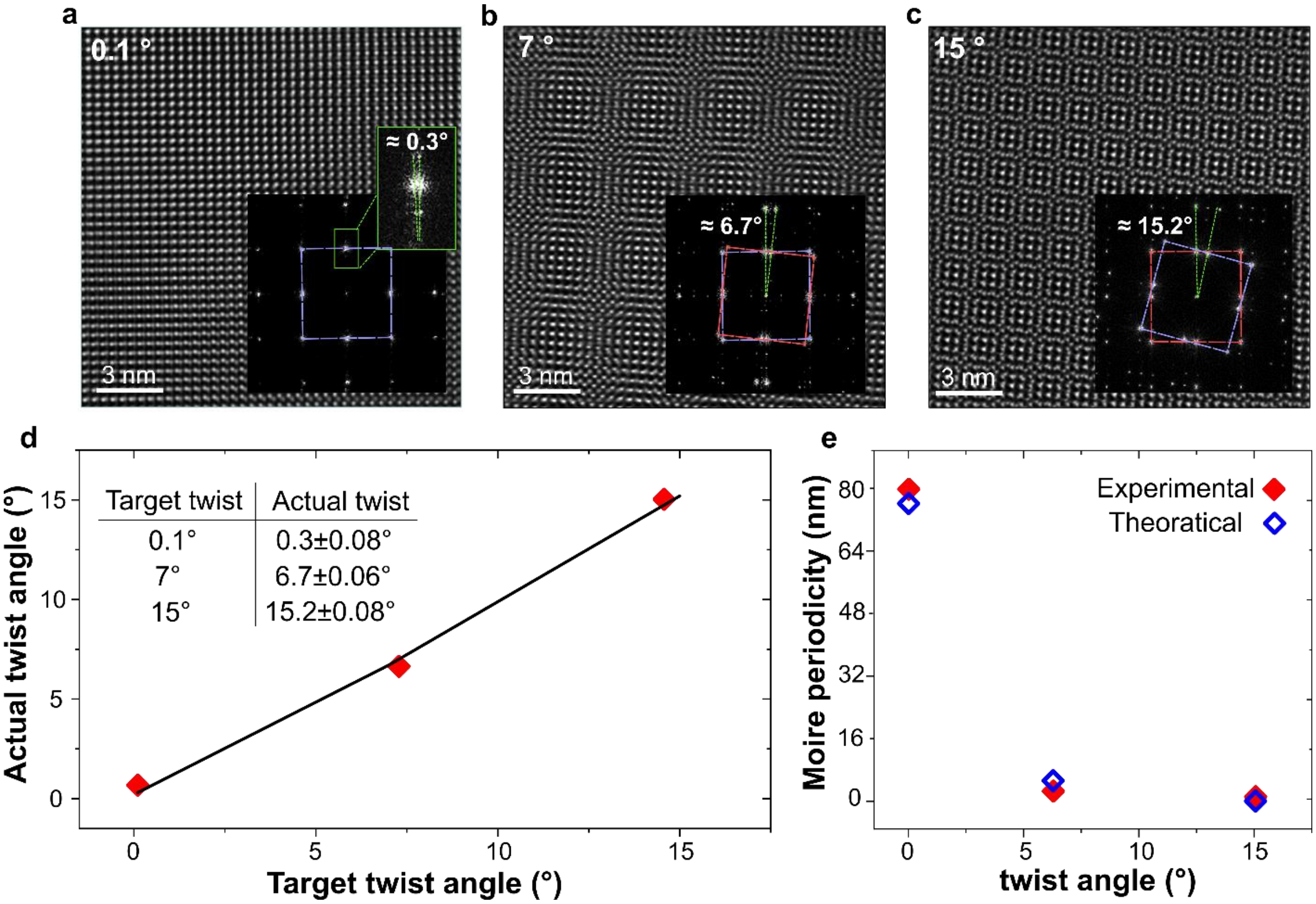

**Figure 2. Atomic-scale verification of twist angles in bilayer oxide membranes.** Atomic-resolution HAADF-STEM images of twisted $NaNbO_3$ membranes with intended twist angles of **(a)** 0.1°, **(b)** 7°, and **(c)** 15°. The actual twist angles were quantitatively extracted from fast Fourier transform (FFT) patterns. **(d)** Comparison between the extracted twist angles from STEM images and the intended twist angles. Error bars, smaller than the marker size, represent the standard deviation. **(e)** Comparison of the experimentally measured and theoretically calculated moiré periodicity with varying twist angles.

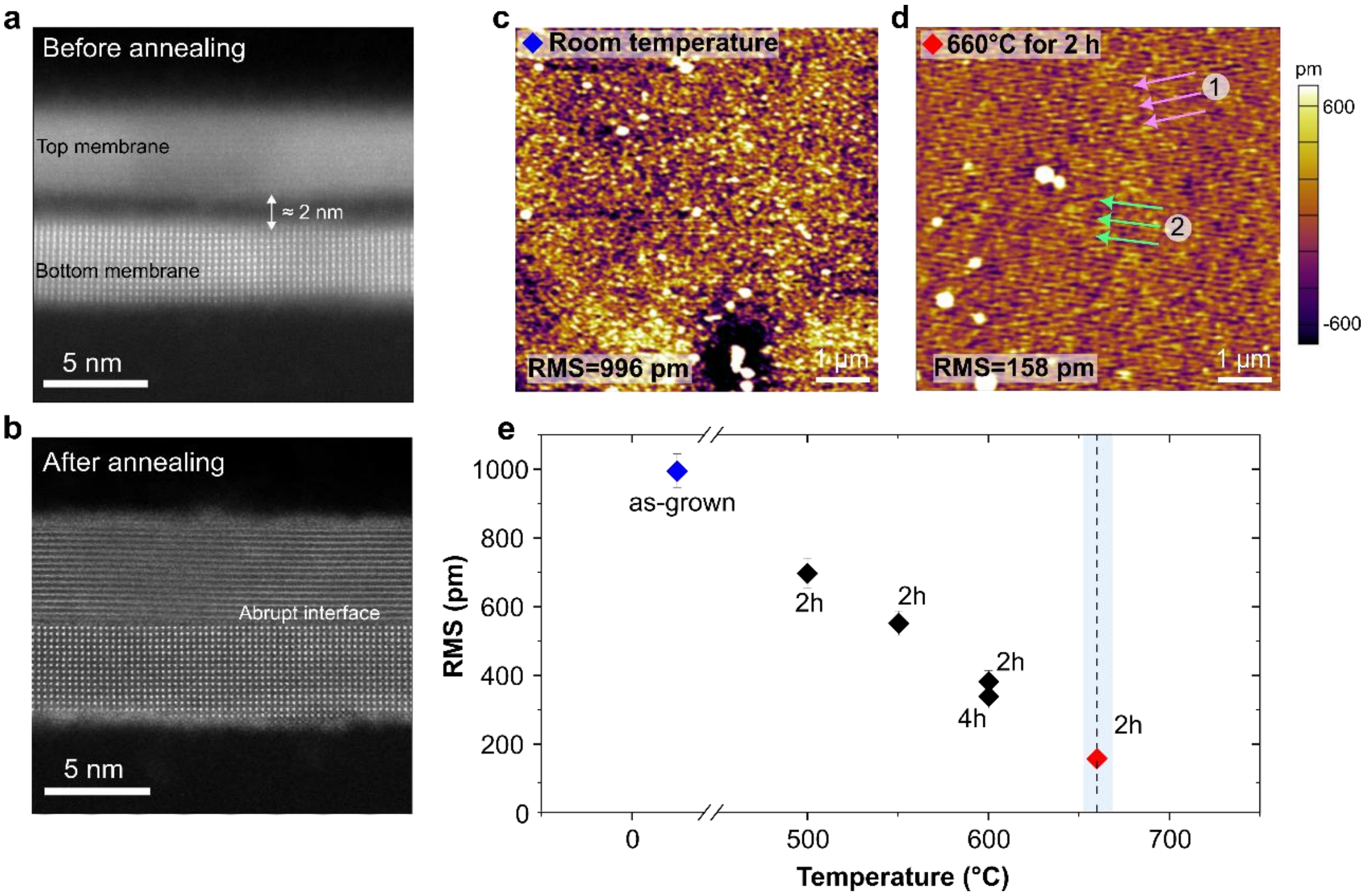

**Figure 3. Effects of thermal annealing on the surface and interfacial roughness of twisted bilayer oxide membranes.** Atomic resolution cross-sectional HAADF-STEM images of the bilayer membranes, **(a)** before and **(b)** after annealing at 660 °C for 2 h, revealing the elimination of the amorphous gap upon annealing and the formation of a clean, atomically coherent interface. Atomic force microscopy (AFM) images of twisted $NaNbO_3$ bilayer membranes collected at **(c)** room temperature prior to annealing and **(d)** after annealing at 660 °C for 2 h. Arrows indicate the step-terrace directions associated with the top and bottom membranes. **(e)** The extracted RMS roughness as a function of annealing temperature. The surface roughness systematically decreases with increasing annealing temperature.

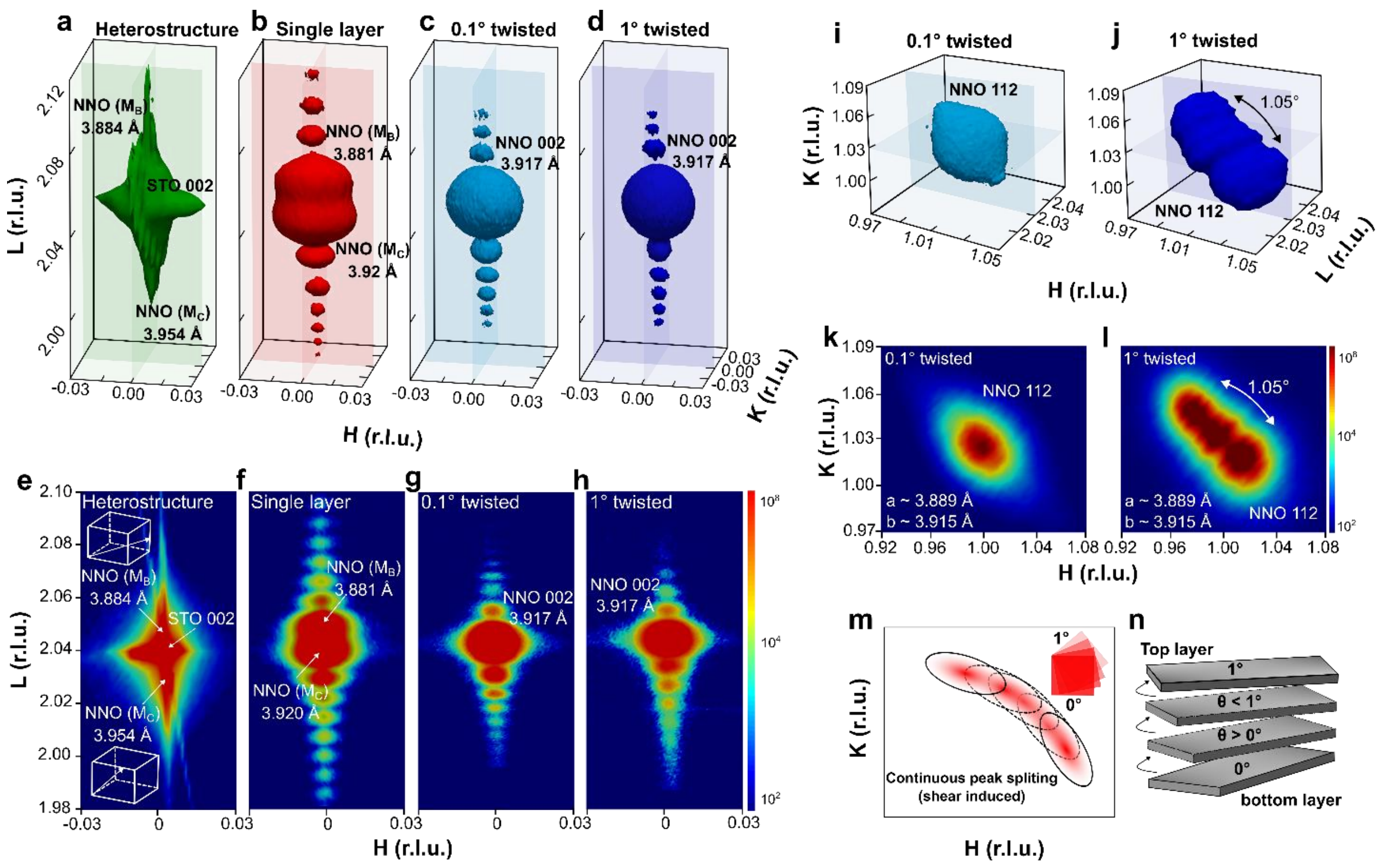

**Figure 4. Structural evolution in $NaNbO_3$ from as-grown heterostructures to single-layer and twisted bilayer membranes.** Synchrotron-based 3D-reciprocal space maps (RSMs) acquired near the pseudocubic 002-diffraction condition, together with the corresponding 2D H-L plane projections for **(a, e)** a 15 nm thick as-grown $NaNbO_3/SrTiO_3$ (001) heterostructure, **(b, f)** a 15 nm thick single-layer $NaNbO_3$ membrane, and a twisted bilayer $NaNbO_3$ membrane with **(c, g)** 0.1° and **(d, h)** 1° twist angles, where each membrane layer is 15 nm thick. Synchrotron-based 3D-RSMs near the pseudocubic 112-diffraction condition and the corresponding 2D H-K plane projections for twisted $NaNbO_3$ bilayer membranes with **(i, k)** 0.1° and **(j, l)** 1° twist angles. **(m, n)** Schematic illustration (exaggerated for clarity) of the continuous in-plane peak splitting observed in (l), indicative of an interlayer-coupling-driven structural reconstruction that accommodates interfacial shear strain through continuous lattice rotations propagating across the heterostructure thickness.

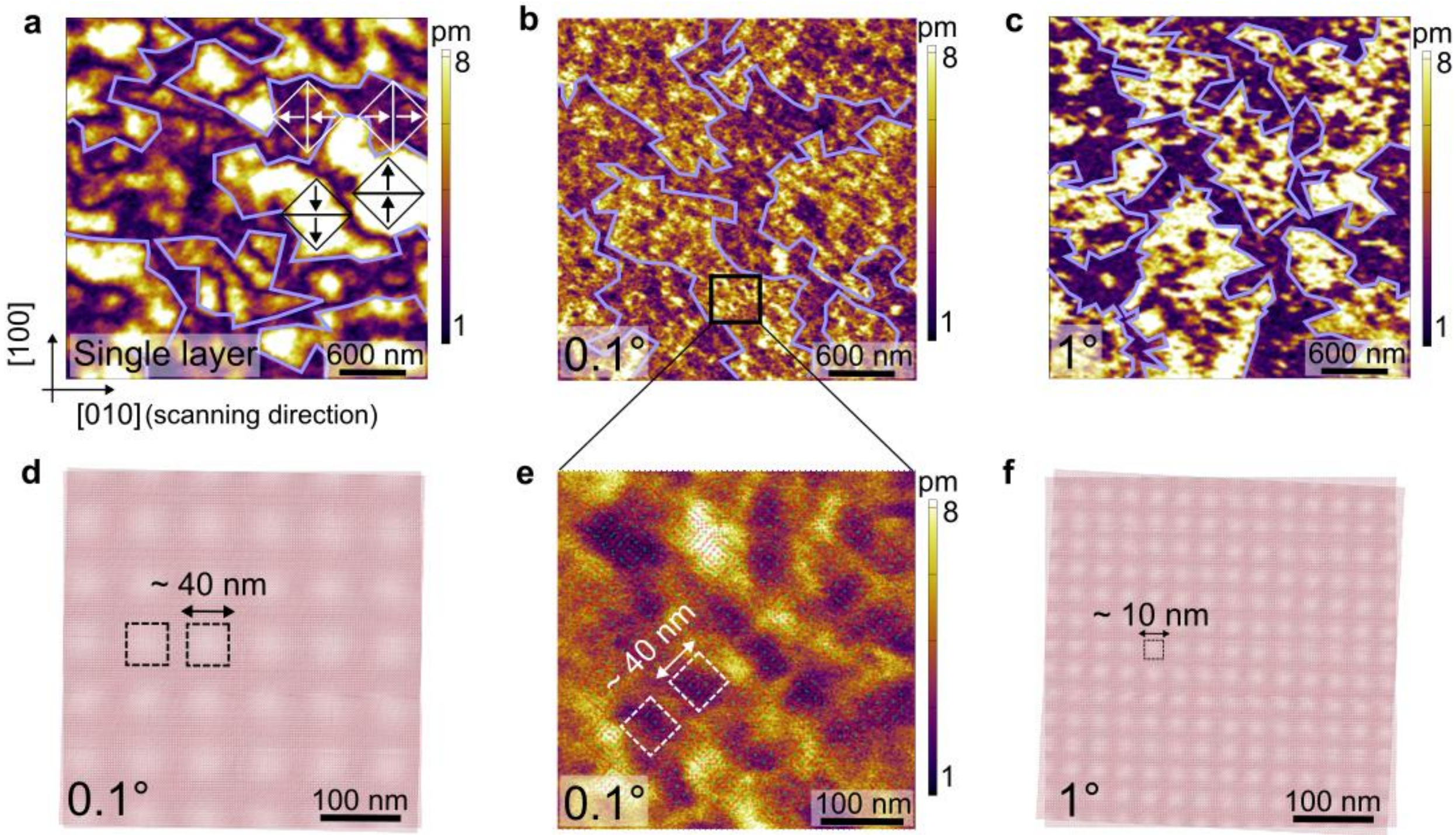

**Figure 5. Evolution of ferroelectric domain structures in $NaNbO_3$ from single-layer to twisted bilayer membranes.** Lateral piezoresponse force microscopy (PFM) amplitude images of **(a)** a single-layer $NaNbO_3$ membrane and twisted bilayer $NaNbO_3$ membranes with intended **(b)** 0.1° and **(c)** 1° twist angles. The arrows in (a) indicate the polarization orientations corresponding to the four possible domain variants of the $M_C$ phase. **(d)** The theoretically predicted moiré superlattice periodicity for 0.1° twisted bilayer membranes. **(e)** Zoomed-in lateral PFM amplitude image of the 0.1° twisted bilayer membranes overlaid with the corresponding predicted moiré superlattice. **(f)** The theoretically predicted moiré superlattice periodicity for 1° twisted bilayer membranes.